\documentstyle[11pt,newpasp,twoside,epsf]{article}
\def\edcomment#1{\iffalse\marginpar{\raggedright\sl#1\/}\else\relax\fi}
\reversemarginpar

\begin{document}
\title{Application of the Variability $\rightarrow$ Luminosity
Indicator to X-Ray Flashes}
\author{D. E. Reichart$^1$, D. Q. Lamb$^2$, R. M. Kippen$^3$, J.
Heise$^4$, J. J. M. in't Zand$^{4,5}$, M. Nysewander$^1$}
\affil{$^1$University of North Carolina, 
$^2$University of Chicago, 
$^3$Los Alamos National Laboratory, 
$^4$Space Research Organization Netherlands, 
$^5$University of Utrecht}
%\author{P. M. Woods}
%\affil{NSSTC}

\begin{abstract}
We have applied the proposed variability $\rightarrow$ luminosity
indicator to ten ``X-Ray Flashes'' (XRFs) observed by the Wide-Field
Cameras on BeppoSAX for which BATSE survey data exists.  Our results
suggest that the variability $\rightarrow$ luminosity indicator
probably works for XRFs.  Assuming this to be so, we find that the
luminosity and redshift distributions of XRFs are consistent with those
of long-duration gamma-ray bursts (GRBs), and therefore most XRFs are
probably not very high redshift GRBs.  The fact that XRFs and GRBs have
similar luminosity and redshift distributions suggests that XRFs and
long-duration GRBs are produced by a similar mechanism.
\end{abstract}

\section{Introduction}

X-ray flashes (XRFs) are a new type of fast transient source observed
with the {\it Beppo}SAX Wide Field Cameras (WFCs) at a rate of about
four per year (Heise et al. 2001).  The rate for XRFs is thus $\approx$
40\% of the rate for gamma-ray bursts (GRBs).

XRFs are distinguished from Galactic transient sources by their
isotropic spatial distribution and short ($\sim$ 10-100 sec)
durations.  Furthermore, they are distinguished from GRBs based on the
fact that the are not detected above 40 keV by the {\it Beppo}SAX GRB
Monitor -- implying much larger $L_x/L_\gamma$ ratios than GRBs.
Actually, the distribution of $L_x/L_\gamma$ ratios overlaps
considerably with that of GRBs (Heise et al. 2001).  In other respects,
such as duration, temporal structure, spectrum and spectral evolution,
XRFs exhibit properties that are qualitatively similar to the X-ray
properties of GRBs.  These similarities have led to the suggestion that
the XRFs are in fact very ``X-ray rich'' GRBs (Heise et al. 2001).

However, no host galaxy nor redshift has yet been reported for an XRF.
Therefore, it is not known whether XRFs are very high redshift GRBs,
normal GRBs viewed off the axis of the jet, or GRBs that have unusually
small energies.

\section{Results}

A possible Cepheid-like luminosity estimator for GRBs was suggested  by
Ramirez-Ruiz and Fenimore (2000) and has been developed further by
Reichart et al. (2000; see also Reichart \& Lamb 2001a,b).  These
authors have shown that there exists a correlation between a measure
$V$ of the variability of the burst time history and the intrinsic
isotropic-equivalent peak photon energy luminosity $L$ of the burst for
the 19 GRBs for which some redshift information (either a redshift
measurement or a redshift limit) exists.  We show this correlation in
Figure 1.

\begin{figure}[t]
\plotfiddle{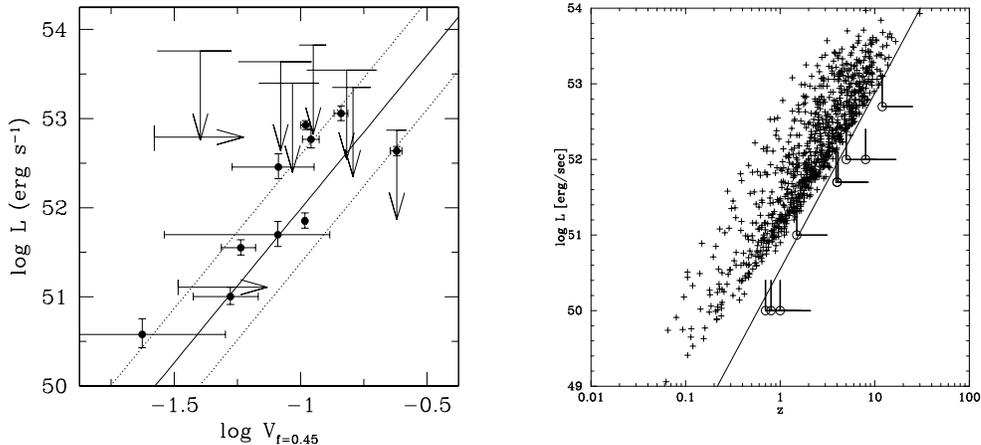}{5.5cm}{0.0}{32.0}{32.0}{-200}{-66}
\plotfiddle{lambd1_2.ps}{0.0cm}{270.0}{33.0}{33.0}{0}{205}

\caption{{\it Left:} The variabilities $V$ and isotropic-equivalent
peak photon energy luminosities $L$ of the 19 bursts for which some
redshift information exists.  The solid and dotted lines mark the
center and 1-$\sigma$ widths of the best-fit model of these bursts in
the (log $L$, log $V$)-plane.
{\it Right:} Comparison of the distribution in the ($z,L$)-plane of the
lower limits in luminosity and redshift for the ten XRFs (open circles)
with the distribution in the ($z,L$)-plane of the luminosities and
redshifts for $\approx 900$ GRBs (crosses) from the BATSE 4B
catalog estimated (Reichart \& Lamb 2001a,b) using the variability
measure of Reichart et al. (2001). }

\end{figure}

In this paper, we use the possible Cepheid-like luminosity estimator
for the long-duration GRBs developed by Reichart et al. (2001) to
estimate the intrinsic isotropic-equivalent peak photon energy
luminosity $L$, and thus the redshift $z$, of 10 XRFs detected by the
{\it Beppo}SAX WFCs (Heise et al. 2001) and the Burst and Transient
Source Experiment (BATSE) on the {\it Compton Gamma-Ray Observatory}
(Kippen et al. 2002).  There is a significant limitation to what we can
say, since these XRFs did not trigger BATSE and therefore we must use
BATSE survey data for these ten events.  The BATSE survey data is 1024
msec time resolution data, whereas all previous applications of our
proposed variability measure have used 64 msec time resolution data.
This means that, technically, we can only establish lower limits on the
luminosities, and therefore the redshifts, of the ten XRFs.  However,
our experience applying our variability measure to $\approx 900$ GRBs
in the BATSE 4B catalog (Paciesas et al. 1999) suggests that this may
not be a severe limitation.

\begin{table}[h]
\begin{center}
\begin{tabular}{llll}
\hline\hline
Burst	& Variability	& Luminosity	& Redshift	\\ 
		&				& \ \ [erg s$^{-1}$] & \\
\hline
971019	& $> 0.15$      & $> 5 \times 10^{52}$	& $> 12$ \\
971024	& $> 0.1$       & $> 1 \times 10^{52}$	& $> 5$ \\
980128	& $> 0.025$     & $> 1 \times 10^{50}$	& $> 1.0$ \\
980306	& $> 0.025$     & $> 1 \times 10^{50}$	& $> 0.8$ \\
990520	& $> 0.075$     & $> 5 \times 10^{51}$	& $> 4$ \\
990526	& $> 0.075$     & $> 5 \times 10^{51}$	& $> 4$ \\
991106	& $> 0$         & $> 0$         		& $> 0$ \\
000206	& $> 0.025$     & $> 1\times 10^{50}$	& $> 0.7$ \\
000208	& $> 0.05$      & $> 1\times 10^{51}$	& $> 1.5$ \\
000416	& $> 0.1$       & $> 1\times 10^{52}$	& $> 8$ \\
\hline\hline
\end{tabular}
\caption{Lower Limits on the Variability, Luminosity, and Redshift of Ten XRFs.}
\end{center}
\end{table}

Table 1 lists our results.  Figure 2 compares the distribution in the
($z,L$)-plane of the lower limits in luminosity $L$ and redshift $z$
for the ten XRFs with the distribution in the ($z,L$)-plane of the
luminosities and redshifts of $\approx 900$ GRBs from the BATSE 4B
catalog (Paciesas et al. 1999) estimated (Reichart \& Lamb 2001a,b)
using the variability measure $V$ of Reichart et al. (2001).  The
diagonal solid line shows the approximate BATSE detection threshold.
Figure 2 shows that the distribution of the lower limits for the XRFs
is indistinguishable from the distribution of the estimates for the
GRBs.

The fact that all ten XRFs lie below but near the BATSE detection
threshold is expected because none of the ten events triggered the
BATSE instrument.   This fact also indicates that the variability
measures $V$ (and therefore the estimates of the luminosities $L$  and
redshifts $z$) which we would obtain for the ten XRFs, using 64 msec
time resolution data, would not differ much from the lower bounds which
we have obtained using BATSE survey data, which has 1024 msec time
resolution.  Otherwise, the ten XRFs would lie above the BATSE
detection threshold, contradicting the fact that they did not trigger
the BATSE instrument.

These results suggest that the variability $\rightarrow$ luminosity
indicator probably works for XRFs, and that the lower bounds on the
luminosities $L$ and redshifts $z$ that we have obtained for the ten
XRFs do not differ much from the estimates of these quantities that we
would find using higher time resolution data.

Figure 3 compares the cumulative distribution of the lower limits on
$z$ for the XRFs with the cumulative distribution of the estimates of
$z$ for the GRBs.  The two distributions appear similar, and a KS test
confirms that there is no significant difference between them.

\begin{figure}[htb]
\plotfiddle{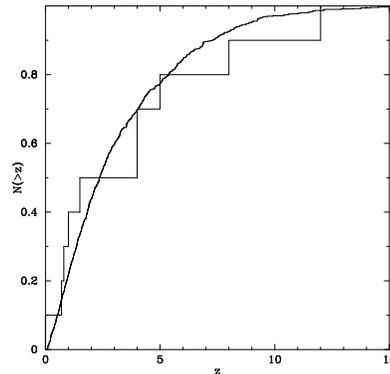}{4.0cm}{270.0}{30.0}{30.0}{-100}{150}

\caption{Comparison of the cumulative distribution of the lower limits
on $z$ for the 10 XRFs with the cumulative distribution of the estimates
of $z$ for the GRBs.  The two distributions appear similar, and a KS
test confirms that there is no significant difference between
them.}\end{figure}

\section{Conclusions} 

Our results suggest that the variability $\rightarrow$ luminosity
indicator probably works for XRFs.  Assuming this to be so, we find
that the luminosity and redshift distributions of XRFs are consistent
with those of long-duration gamma-ray bursts (GRBs), and therefore most
XRFs are probably not very high redshift GRBs.  The fact that XRFs and
GRBs have similar luminosity and redshift distributions suggests that
XRFs and long-duration GRBs are produced by a similar mechanism.


\begin{references}

\reference Heise, J., in't Zand, J. J. M., Kippen, R. M., and Woods, P. M.
2001, in Proceedings of the Rome Workshop on Gamma-Ray Bursts in  the Afterglow
Era, ed. E. Costa, F. Frontera \& J. Hjorth, 16

\reference Kippen, R. M., Woods, P. M., Heise, J., in't Zand, J. J. M., Briggs,
M.S. and Preece, R. D. 2002, in Proceedings of the Woods Hole Workshop on
Gamma-Ray Bursts, ed. R. Vanderspek \&  G. R. Ricker (New York: AIP), in press

\reference Paciesas, W. S., et al. 1999, ApJS, 122, 465

\reference Ramirez-Ruiz, E. \& Fenimore, E. E. 2000, ApJ, submitted
(astro-ph/0004176)

\reference Reichart, D.,  Lamb, D., Fenimore, E., Ramirez-Ruiz, E., Cline, T.
\& Hurley, K.,  et al. 2001, ApJ, 552, 57

\reference Reichart, D. E. and Lamb, D. Q. 2001a, in Gamma-Ray Bursts in  the
Afterglow Era, ed. E. Costa, F. Frontera, and J. Hjorth (Springer-Verlag;
Berlin), 233

\reference Reichart, D. E. and Lamb, D. Q.  2001b, in Relativistic
Astrophysics, AIP Conference Proceedings No. 586, ed. J. C. Wheeler and  H.
Martel (AIP: New York), p. 599

\end{references}
\end{document}